\documentclass[journal]{IEEEtran}

\usepackage[numbers, sort&compress]{natbib}

\usepackage{amsmath, amssymb, amsfonts}
\usepackage{empheq}
\usepackage{booktabs}
\usepackage{threeparttable}
\usepackage{multirow}
\usepackage{multicol}

\usepackage[thinlines, thiklines]{easybmat}

\usepackage[linesnumbered, ruled, vlined]{algorithm2e}

\usepackage{graphicx, subfigure, svg}  

\usepackage[english]{babel}

\usepackage{arydshln}

\usepackage{bbm}

\usepackage{enumerate}
\usepackage{paralist}
\usepackage{url}
\usepackage{color}

\usepackage{url}
\usepackage{hyperref}

\begin{document}

\renewcommand{\figurename}{Fig.}
\renewcommand{\tablename}{TABLE}

\title{Open-Set RF Fingerprinting \\ via Improved Prototype Learning}


\author{
    \IEEEauthorblockN{Weidong~Wang, Hongshu~Liao, and Lu~Gan}
    \thanks{W. Wang, H. Liao, and L. Gan are with the School of Information and Communication Engineering, University of Electronic Science and Technology of China, Chengdu 611731, China. (e-mail: wwdong@std.uestc.edu.cn; hsliao@uestc.edu.cn; ganlu@uestc.edu.cn).}
}

\markboth{Communication Signal Recognition}
{Wang \MakeLowercase{\textit{et al.}}: Open-Set RF Fingerprinting}

\maketitle

\begin{abstract}
    Deep learning has been widely used in radio frequency (RF) fingerprinting. Despite its excellent performance, most existing methods only consider a closed-set assumption, which cannot effectively tackle signals emitted from those unknown devices that have never been seen during training. In this letter, we exploit prototype learning for open-set RF fingerprinting and propose two improvements, including consistency-based regularization and online label smoothing, which aim to learn a more robust feature space. Experimental results on a real-world RF dataset demonstrate that our proposed measures can significantly improve prototype learning to achieve promising open-set recognition performance for RF fingerprinting.
\end{abstract}

\begin{IEEEkeywords}
    RF fingerprinting,
    open-set recognition,
    prototype learning,
    consistency-based regularization,
    online label smoothing.
\end{IEEEkeywords}

\IEEEpeerreviewmaketitle

\section{Introduction} \label{Section: Introduction}
\IEEEPARstart{I}{n} recent years, deep learning has achieved great success in computer vision and natural language processing \cite{lecun2015deep, pouyanfar2018survey}, which provides new insights into many wireless communication issues, including radio frequency (RF) fingerprinting \cite{soltanieh2020review}, a promising physical layer authentication technique that aims to distinguish different transmitters by characterizing device-specific features (aka ``fingerprints") presented in their emitted signals. The powerful non-linear representation of deep neural networks makes it possible to learn high-level features directly from raw communication signal data with complex structures and inner correlations, which brings better performance and many other potential advantages. Nonetheless, RF fingerprinting based on deep learning still suffers from many pressing challenges in real-world applications.

A typical challenge is that most existing methods \cite{riyaz2018deep, yu2019robust, sankhe2019no} only consider an ideal closed-set assumption that all testing classes are known at training time. The broadcast nature of wireless communication makes air interfaces not only open and accessible to authorized users \cite{zou2016survey}. When encountering signals from unknown devices, conventional deep neural networks will mistakenly identify them as one of these knowns. Open-set recognition (OSR) is a promising solution to overcome this issue, where an open-set classifier should be able to reject samples of unknown classes and maintain high classification accuracy on knowns.

The concept of open-set recognition was originally defined theoretically by Scheirer \emph{et al.} \cite{scheirer2012toward}, who added a hyperplane in support vector machines (SVM) to distinguish samples of known classes from unknowns. Due to such impressive performance of deep neural networks in a wide range of classification tasks, open-set recognition based on deep learning has attracted extensive attention. Bendale and Boult proposed OpenMax \cite{bendale2016towards}, which exploits extreme value theory to reallocate logits for constructing an explicit probability of unknown classes. Ge \emph{et al.} \cite{ge2017generative} and Neal \emph{et al.} \cite{neal2018open} attempted to use generative adversarial networks (GAN) to synthesize samples of unknown classes for training an open-set classifier. Furthermore, many reconstruction-based approaches \cite{yoshihashi2019classification, sun2020conditional} have been widely studied, among which Sun \emph{et al.} \cite{sun2020conditional} achieved a promising result by learning a conditional Gaussian distribution for known classes and detecting unknown ones. Recently, since Yang \emph{et al.} \cite{yang2018robust, yang2020convolutional} attempted to combine prototype learning with deep neural networks for open-set recognition, prototype learning-based approaches \cite{chen2021adversarial, lu2022pmal} have reached state-of-the-art (SOTA) performance.

\begin{figure}[htb]
    \centering
    \subfigure[Training]
    {
        \centering
        \includegraphics[width=0.22\textwidth]{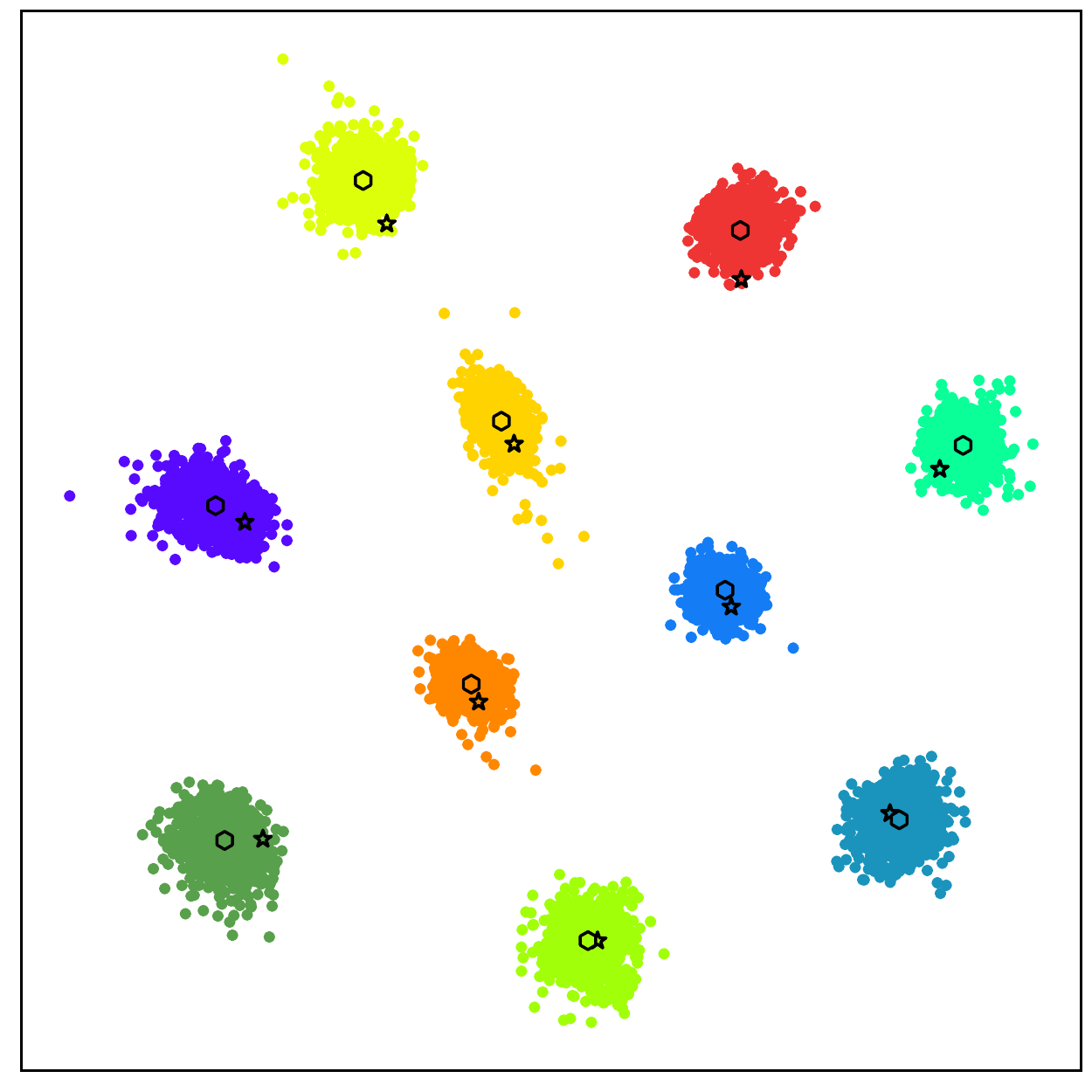}
    }
    \hspace{0.1em}
    \subfigure[Validation]
    {
        \centering
        \includegraphics[width=0.22\textwidth]{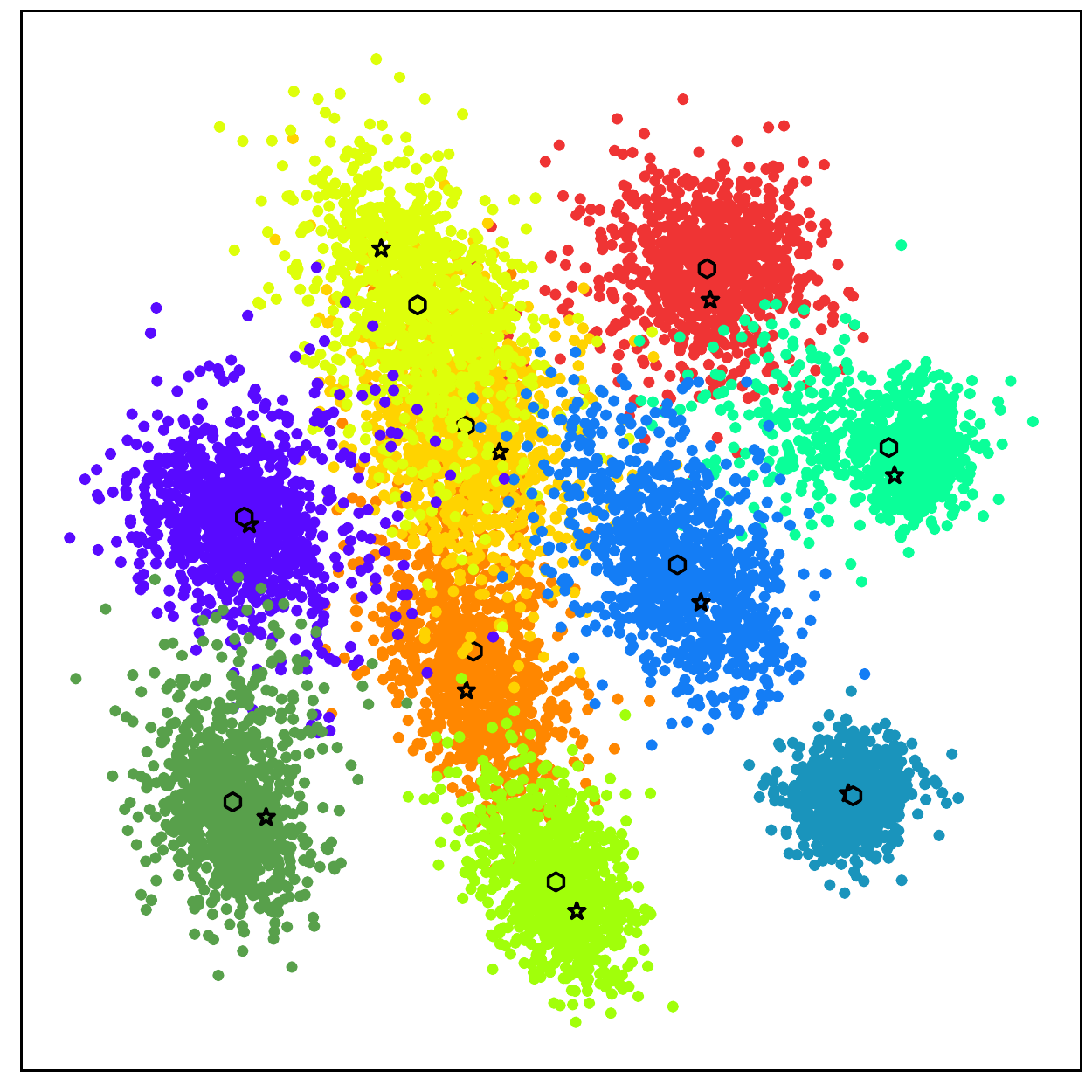}
    }
    \caption{Visualization of 2-D embedded feature space learned by GCPL. The cluster center of each class and its corresponding prototype are highlighted using a hexagon marker and a star marker, respectively.}
    \label{Figure: Feature Space}
\end{figure}

Prototypes are those representative samples or potential feature representations of each class. In generalized convolutional prototype learning (GCPL) \cite{yang2018robust, yang2020convolutional}, Yang \emph{et al.} proposed a regularization form termed ``prototype loss", which allows deep neural networks not to make a partition for the whole feature space but only project samples onto specific regions centered around each prototype. The prototype loss actually represents the average radius of the prototype clusters. Theoretically, as long as this radius is small enough, ideally to be zero, GCPL can detect any unknown classes because the feature representations of unknowns cannot be exactly coincident with knowns. In practice, however, such prototype clusters on validation samples are often far less compact than on training samples, especially for relatively complex data (e.g., communication signal data with very small RF fingerprints \cite{soltanieh2020review}), as shown in Fig. \ref{Figure: Feature Space}. This phenomenon indicates that GCPL may have certain overfitting. The feature space it learned is still not robust enough.

To address this issue, we propose an improved version of GCPL, called \textbf{I}mproved \textbf{P}rototype \textbf{L}earning (IPL). In contrast to GCPL, we introduce data augmentation and perform consistency-based regularization on feature representations, which requires a sample and its augmented version to have very close feature representations, thus enforcing deep neural networks to learn a more robust feature space. Moreover, since label smoothing can enlarge inter-class distances and reduce intra-class differences \cite{muller2019does, shen2021label}, we carefully design an online label smoothing technique, which can further promote learning a better feature space. Experimental results on a real-world RF dataset, WIDEFT \cite{siddik2021wideft}, demonstrate that our proposed measures can significantly improve prototype learning to achieve promising open-set recognition performance for RF fingerprinting.

\section{Methodology} \label{Section: Methodology}

\subsection{Preliminaries}
Before introducing related improvements, we briefly review generalized convolutional prototype learning (GCPL) \cite{yang2018robust, yang2020convolutional}, which learns and maintains several prototypes on feature representations of each class and uses prototype matching for classification. It is important to note that an ideal distribution of feature representations should be nearly Gaussian. Moreover, we have observed that GCPL usually learns only one effective prototype for each class, even when configured with multiple prototypes, which may be attributed to cross-entropy. For clarity, we only describe its single prototype form.

\begin{figure}[htb]
    \centering
    \includegraphics[scale=0.72]{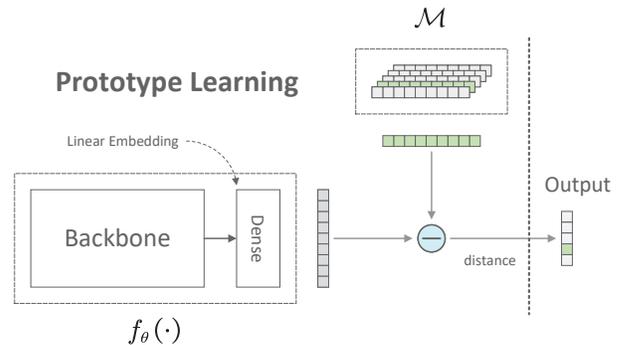}
    \caption{Illustration of generalized prototype learning.}
    \label{Figure: Generalized Prototype Learning}
\end{figure}

The network architecture of GCPL consists of a convolutional feature extractor $f_\theta(\cdot)$ and a group of learnable prototypes $\mathcal{M}$, as illustrated in Fig. \ref{Figure: Generalized Prototype Learning}. The prototypes are denoted as $m_{k} \in \mathcal{M}$, where $k \in \{1, 2, \ldots, \mathrm{C}\}$ if there are $\mathrm{C}$ known classes. Formally, we are provided with a dataset $\mathcal{D} = \{(\boldsymbol{x}, \, y)\}$ to train $f_\theta(\cdot)$ and $\mathcal{M}$ jointly in an end-to-end manner. The loss function of GCPL is defined as
\begin{equation}
    \mathcal{L} = \mathcal{L}_{\mathrm{DCE}} + \lambda \mathcal{L}_{\mathrm{prototype}}
\end{equation}
where $\mathcal{L}_{\mathrm{DCE}}$ is a classification loss termed ``distance-based cross-entropy", $\mathcal{L}_{\mathrm{prototype}}$ represents a regularization form termed ``prototype loss", and $\lambda$ is a small penalty factor.

The distance-based cross-entropy $\mathcal{L}_{\mathrm{DCE}}$ is given by
\begin{equation}
    \begin{split}
        \mathcal{L}_{\mathrm{DCE}}(\boldsymbol{x}, \, y)
        &= - \log p\left(y \mid \boldsymbol{x}\right) \\
        &= - \log p\left(f_{\theta}(\boldsymbol{x}) \in m_{y} \mid \boldsymbol{x} \right)
    \end{split}
\end{equation}
where
\begin{equation}
    p\left(f_{\theta}(\boldsymbol{x}) \in m_{k} \mid \boldsymbol{x} \right) =
    \frac{e^{-\gamma \operatorname{distance}(f_{\theta}(\boldsymbol{x}), \, m_{k})}}{\sum\limits_{i=1}^{\mathrm{C}} {e^{-\gamma \operatorname{distance}(f_{\theta}(\boldsymbol{x}), \, m_{i})}}}
\end{equation}
is a negative log-likelihood to transform distance metric space to probability space, we also write it as $p_{k}(\boldsymbol{x})$ for short, and $\gamma$ is a hyper-parameter that control the hardness of probability conversion. The function $\operatorname{distance}(\cdot)$ is used to compute the Euclidean distance, i.e.,
\begin{equation}
    \operatorname{distance}(f_{\theta}(\boldsymbol{x}), \, m_{k}) = \left\Vert f_{\theta}(\boldsymbol{x}) - m_{k} \right\Vert^2
\end{equation}

The prototype loss $\mathcal{L}_{\mathrm{prototype}}$ is defined as
\begin{equation}
    \mathcal{L}_{\mathrm{prototype}}(\boldsymbol{x}, \, y) = \left\Vert f_{\theta}(\boldsymbol{x}) - m_{y} \right\Vert^2
\end{equation}
which can pull the feature representations of samples close to their corresponding prototypes, making those feature representations within an identical class more compact. This regularization also avoids to learn a linear decision boundary like softmax that linearly separates the whole feature space.

\subsection{Learning of More Robust Feature Space}
To learn a more robust feature space, we propose two effective improvements: consistency-based regularization and online label smoothing. The details are as follows.

\vspace{10pt}

\noindent\textbf{Consistency-Based Regularization} The effect of prototype learning, i.e., compactness of prototype clusters, is significantly different between training and validation data, indicating a lack of generalization. The learned feature space is still not robust enough. To this end, we introduce data augmentation and perform consistency-based regularization on feature representations, which requires a sample and its augmented version to have very close feature representations, i.e.,
\begin{equation}
    \mathcal{L}_{\mathrm{consistency}}(\boldsymbol{x}) = \left\Vert f_{\theta}(\boldsymbol{x}) - f_{\theta}(\operatorname{g}(\boldsymbol{x})) \right\Vert^2
\end{equation}
where $\operatorname{g}(\cdot)$ represents a data augmentation operation. It is recommended to only use data augmentations that do not change sample distribution and have certain randomness. The total objective thus can be rewritten as
\begin{equation}
    \mathcal{L} = \mathcal{L}_{\mathrm{DCE}} + \lambda_1 \mathcal{L}_{\mathrm{prototype}} + \lambda_2 \mathcal{L}_{\mathrm{consistency}}
\end{equation}
Note that consistency-based regularization is not a new concept, and it has been widely used in deep semi-supervised learning \cite{sohn2020fixmatch}, where an unlabeled sample and its augmented version must have similar model predictions. It can be seen that ours is more strict, and we perform this constraint on feature representations rather than model predictions. This essentially belongs to a self-supervised mechanism.

\vspace{10pt}

\noindent\textbf{Online Label Smoothing} As a commonly used regularization trick, label smoothing \cite{szegedy2016rethinking} can prevent deep neural networks from becoming over-confident and thus improves generalization to a certain extent. More recently, Müller \emph{et al.} \cite{muller2019does} found that label smoothing tends to erase information contained intra-class across individual samples, enforcing each sample to be equidistant to its template. The empirical study of Shen \emph{et al.} \cite{shen2021label} further pointed out that this ``erasing" effect enabled by label smoothing actually enlarges relative information on those semantically similar classes, i.e., making them have less overlap on feature representations. These two properties undoubtedly benefit prototype learning and can promote learning a better feature space.

\begin{figure}[htb]
    \centering
    \includegraphics[scale=0.65]{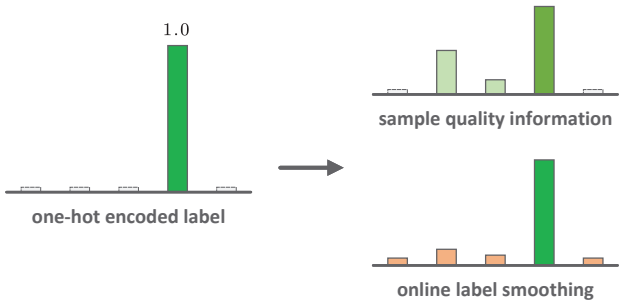}
    \caption{Illustration of online label smoothing.}
    \label{Figure: Online Label Smoothing}
\end{figure}

Label smoothing replaces a one-hot encoded label with a mixture of its weighted value and a uniform distribution, i.e.,
\begin{equation}
    \tilde{y}_{k} =
    \begin{cases}
        1 - \alpha                                         & \text{if} \ k = y \ \text{(i.e., target class),} \\
        \displaystyle \alpha / \left(\mathrm{C} - 1\right) & \text{otherwise.}
    \end{cases}
\end{equation}
where $\alpha$ is a small constant coefficient to flatten one-hot encoded labels. The classification loss $\mathcal{L}_{\mathrm{DCE}}$ with such a softened label can be rewritten as
\begin{equation}
    \mathcal{L}_{\mathrm{DCE}}(\boldsymbol{x}, \, \tilde{\boldsymbol{y}}) =
    = - \sum_{k}^{\mathrm{C}} \tilde{y}_k \log p_{k}\left(\boldsymbol{x}\right)
\end{equation}

To achieve a better ``erasing" effect, we propose online label smoothing, as illustrated in Fig. \ref{Figure: Online Label Smoothing}, and it enjoys more label semantics than conventional label smoothing. Given a sample $\boldsymbol{x}$ and its label $y$, we maintain a cumulative record of model predictions, denoted by $\boldsymbol{q}$, which is initialized as a one-hot according to $y$ and then persistently updated during training. Let $\boldsymbol{q}$ of epoch $t$ be $\boldsymbol{q}^{(t)}$, and we update it at each epoch ending, i.e.,
\begin{equation}
    \boldsymbol{q}^{(t)} \leftarrow \boldsymbol{q}^{(t)} + \boldsymbol{p}^{(t)}
\end{equation}
where $\boldsymbol{p}^{(t)}$ represents the corresponding model prediction for $\boldsymbol{x}$ at epoch $t$. It is not difficult to find that $\boldsymbol{q}$ can reflect a relative relation between the target class $y$ and other classes, like which class this sample is easily misclassified. Hence, we also refer to $\boldsymbol{q}$ as ``sample quality information". The proposed online label smoothing exploits such sample quality information with a carefully designed rule to build a softened label during training, as shown below.
\begin{equation}
    \tilde{y}_{k} = \displaystyle p_{\upsilon} + (p_{\max} - p_{\min}) \frac{q_{k}}{\varsigma} \quad \text{for} \ k = 1, \, 2, \, \ldots, \, \mathrm{C}
\end{equation}
where $p_{\min} = 0.6$, $p_{\max} = 1 - \alpha$, $\varsigma = \sum_{i} q_{i}$, and
\begin{equation}
    p_{\upsilon} =
    \begin{cases}
        0.6                                                & \text{if} \ k = y \ \text{(i.e., target class),} \\
        \displaystyle \alpha / \left(\mathrm{C} - 1\right) & \text{otherwise.}
    \end{cases}
\end{equation}




\subsection{Rejection Rule of Unknowns}
Generally, a rejection rule based on posterior probability is used to detect ambiguous patterns (those belonging to known classes with low confidence) in closed-set recognition \cite{dubuisson1993statistical}. Given a threshold $\tau$, if
\begin{equation}
    \mathop{\max}_{k}^{\mathrm{C}} p_{k}(\boldsymbol{x}) < \tau
\end{equation}
then $\boldsymbol{x}$ is identified as unknown, otherwise it is determined as known. This rejection rule can be derived into another more efficient form in prototype learning \cite{yang2020convolutional}, as follows.
\begin{equation}
    g_{i1}(\boldsymbol{x}) - g_{i2}(\boldsymbol{x}) < \tau
\end{equation}
where $i1$ and $i2$ denote the classes which match best and second-best with the sample, and
\begin{equation}
    g_{i}(\boldsymbol{x}) = - \left\Vert f_{\theta}(\boldsymbol{x}) - m_{i} \right\Vert^2
\end{equation}

Note that we still use this rejection rule for open-set recognition, but present a practical solution for adaptively setting $\tau$ by considering $\hat{\tau} = g_{i1}(\boldsymbol{x}) - g_{i2}(\boldsymbol{x})$, and then calculating its mean and standard deviation over those correctly classified samples of validation data for each known class, denoted by $\mu$ and $\sigma$, respectively. Let $\tau = \mu - \kappa \sigma$, where $\kappa \ge 1.0$. Then, we can configure $\kappa = 1.0$, i.e., $\tau = \mu - \sigma$ as a relatively optimal threshold, or it is possible to increase $\kappa$ appropriately to obtain higher classification accuracy on knowns.

\section{Experiments and Results} \label{Section: Experiments and Results}
In this section, a series of experiments are conducted to evaluate our proposed method comprehensively.

\subsection{Data Preparation}
An open-source RF dataset, WIDEFT \cite{siddik2021wideft}, is considered for our experiments, available on \href{https://zenodo.org/record/4116383}{Zenodo}. This dataset is collected from $138$ real-world devices (e.g., smartphones, headsets, routers), spanning $79$ unique models, some of which (e.g., smartphones) are capable of more than one wireless type (e.g., Bluetooth/WiFi) and radio modes (e.g., $2.4 \text{/} 5$ GHz WiFi), leading to a total of $147$ signal data captures made. Each capture contains a set of $100$ bursts, each of which is complete that consists of ON transient, steady-state portion, and OFF transient, and includes $5000$ sampling points before and after. The steady-state portion is long enough to be sliced into multiple samples. For open-set recognition, we select $18$ Apple Inc's $2.4$ GHz WiFi devices, among which $10$ are used as known classes, which are A1241, A1349, A1367-1, A1367-2, A1367-3, A1387, A1432, A1453, A1534, A1660. The rest are taken as unknowns. Each burst is randomly sliced into $50$ samples. The signal data of known devices are divided into a training set, validation set, and test set in a proportion of $3:1:1$, while for unknowns, all are only used in testing.

\subsection{Implementation and Training Details}
The backbone used for this work is a well-designed deep residual network (ResNet) \cite{he2016deep} with an initial convolutional layer of $128$ kernels of size $15$, which has been widely used in our other works and can achieve excellent supervised performance for communication signal recognition \cite{wang2023semi}. The dimension size of linear embedding is set to $128$. Also, in our past work \cite{wang2023semi}, a composite operation of rotation and stochastic permutation as data augmentation has been demonstrated to be very effective for RF fingerprinting, which does not change sample distribution and has certain randomness. The model is built with TensorFlow \cite{tensorflow2015} and then trained on a single NVIDIA RTX 2080S GPU utilizing an Adam \cite{kingma2014adam} optimizer for $230$ epochs. The batch size is set to $128$. The initial learning rate is set to $0.001$. For each experimental setting (e.g., ablation study of different hyper-parameter settings), we show its best performance in $10$ consecutive trials.

\subsection{Ablation Study of Improvements}
\begin{figure}[htb]
    \centering
    \includegraphics[scale=0.6]{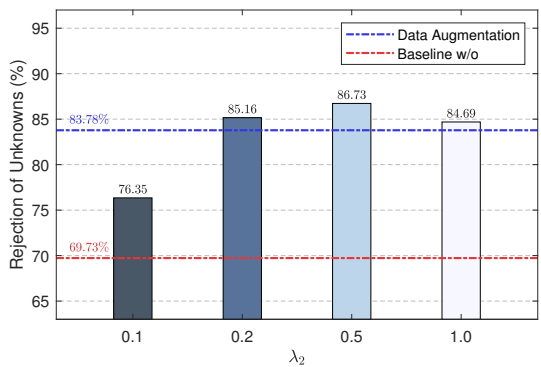}
    \caption{Ablation study of consistency-based regularization.}
    \label{Figure: Ablation Study of Consistency-Based Regularization}
\end{figure}

Fig. \ref{Figure: Ablation Study of Consistency-Based Regularization} compares GCPL improved by consistency-based regularization under different $\lambda_2$ settings. The original version of GCPL is used as a baseline, with $\gamma = 1.0$ and $\lambda = 0.1$. The models are all optimized to have an overall recognition accuracy of $90\%$ on known classes by adjusting their detection thresholds, and then we compare their average rejection rates in different settings. It can be seen that consistency-based regularization significantly improves GCPL and also performs slightly better than purely using data augmentation in most cases. The best performance is achieved at $\lambda_2 = 0.5$.

\begin{figure}[htb]
    \centering
    \includegraphics[scale=0.6]{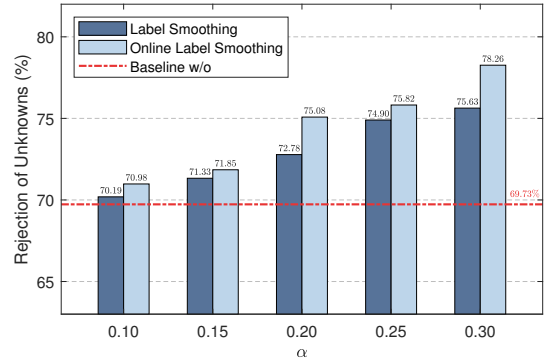}
    \caption{Ablation study of label smoothing.}
    \label{Figure: Ablation Study of Label Smoothing}
\end{figure}

Fig. \ref{Figure: Ablation Study of Label Smoothing} compares label smoothing and our proposed online label smoothing, investigating their improvements to GCPL under different $\alpha$ settings. It can be seen that online label smoothing is superior to conventional label smoothing. While a bigger $\alpha$ is more likely to improve performance, we have observed that it more easily leads to instability for training, including issues like NaN loss or falling into some bad local minima. After considering various factors, we take $\alpha = 0.2$.

\subsection{Comparsion with Other Methods}
The proposed method (i.e., IPL) is compared with other competing ones, including OpenMax \cite{bendale2016towards}, GCPL \cite{yang2018robust, yang2020convolutional}, and ARPL \cite{chen2021adversarial} (The newly recent SOTA for open-set recognition), as shown in Table \ref{Table: Comparsion with Other Methods}. The same backbone is used for all methods. It can be seen that ours is far superior to all these mentioned methods. The proposed method for open-set RF fingerprinting achieves an average rejection rate of $90.28\%$ for unknowns while maintaining high classification accuracy ($\sim 90\%$), which has great practical value.


\begin{table}[htb]
    \renewcommand\arraystretch{1.15}
    \centering
    \caption{Comparsion with Other Methods}
    \label{Table: Comparsion with Other Methods}
    \resizebox{0.475\textwidth}{!}{
        \begin{threeparttable}
            \begin{tabular}{ccccc}
                \toprule
                \textbf{Method}   & OpenMax   & GCPL                      & ARPL                      & Proposal                  \\
                \midrule
                \specialrule{0em}{0.8pt}{0pt}
                \midrule
                \textbf{Adaptive} & //        & $80.22\% \, / \, 86.38\%$ & $80.59\% \, / \, 87.92\%$ & $84.44\% \, / \, 94.76\%$ \\
                \midrule
                \textbf{Reject.}  & $66.82\%$ & $69.73\%$                 & $77.25\%$                 & $90.28\%$                 \\
                \bottomrule
            \end{tabular}
            \begin{tablenotes}
                \item The adaptive results like ``$84.44\% \, / \, 94.76\%$'' means it obtains an overall recognition accuracy of $84.44\%$ on knowns and has an average rejection rate of $94.76\%$ on unknowns when adaptively configuring $\tau = \mu - \sigma$. Note that this item is ignored in OpenMax since it is not a prototype learning-based approach and has more parameters to adjust.
            \end{tablenotes}
        \end{threeparttable}}
\end{table}

\section{Conclusion} \label{Section: Conclusion}
This letter exploits prototype learning for open-set RF fingerprinting and proposes two effective improvements, including consistency-based regularization and online label smoothing. Consistency-based regularization can encourage a sample and its augmented version to have very close feature representations, enforcing deep neural networks to learn a more robust feature space, while online label smoothing can build softened labels with richer semantics, making learned prototype clusters more compact. The experimental results on a real-world RF dataset demonstrate that our proposed measures can significantly improve prototype learning to achieve promising open-set recognition performance for RF fingerprinting.

\bibliographystyle{IEEEtran}
\bibliography{references}

\end{document}